\documentstyle[11pt,aaspp4,epsfig]{article}
\slugcomment{To appear in The Astrophysical Journal Letters, vol. 512 (1999)}

\begin{document}

\title{Strong Aperiodic X-ray Variability and Quasi-Periodic Oscillation in X-ray Nova XTE J1550-564}
\author{Wei Cui\altaffilmark{1}, Shuang Nan Zhang\altaffilmark{2}, Wan
Chen\altaffilmark{3,4}, and Edward H. Morgan\altaffilmark{1}}

\altaffiltext{1}{Center for Space Research, Massachusetts
Institute of Technology, Cambridge, MA 02139; cui@space.mit.edu;
ehm@space.mit.edu}

\altaffiltext{2}{Department of Physics, University of Alabama
in Huntsville, Huntsville, AL 35899; zhangsn@email.uah.edu}

\altaffiltext{3}{NASA/Goddard Space Flight Center, Code 661,
Greenbelt, MD 20771; chen@milkyway.gsfc.nasa.gov}

\altaffiltext{4}{also Department of Astronomy, University of Maryland,
College Park, MD 20742}

\begin{abstract}
We report the discovery of strong aperiodic X-ray variability and 
quasi-periodic oscillation (QPO) in the X-ray light curves of a new
X-ray Nova, XTE J1550-564, and the evolution of the observed 
temporal properties during the rise of the recent X-ray outburst. 
The power spectral analysis of the first observation 
reveals strong aperiodic X-ray variability of the source ($\sim$28\%), 
as well as the presence of a QPO at $\sim$82 mHz with fractional rms 
amplitude $\sim$14\% over the 2 -- 60 keV energy range. Also apparent 
is the first harmonic of the QPO with the amplitude $\sim$9\%. As the 
X-ray flux increases, the source tends to become less variable, and
the QPO frequency increases rapidly, from 82 mHz to 4 Hz, over the 
flux (2 -- 50 keV) range of 
1.73 -- 5.75 $\times 10^{-8}\mbox{ }ergs\mbox{ }cm^{-2}\mbox{ }s^{-1}$.
The amplitude of the fundamental component of the QPO varies little, 
while that of the harmonic follows a decreasing trend. The fundamental 
component strengthens 
toward high energies, while its harmonic weakens. Initially, the power 
spectrum is roughly flat at low frequencies and turns into a power law 
at high frequencies, with the QPO harmonic sitting roughly at the
break. In later observations, however, the high-frequency portion of 
the continuum can actually be better described by a broken power law 
(as opposed to a simple power law). This effect becomes more apparent 
at higher energies. The overall amplitude of the continuum shows 
a similar energy dependence to that of the fundamental component of 
the QPO. Strong rapid X-ray variability, as well as hard energy
spectrum, makes XTE J1550-564 a good black hole candidate. We compare 
its temporal properties with those of other black hole candidates. 

\end{abstract}


\section{Introduction}
A new X-ray nova designated as XTE J1550-564 was discovered by the 
All-Sky Monitor (ASM) aboard the {\it Rossi X-ray Timing Explorer} 
(RXTE) on September 7, 1998 (Smith 1998). A possible optical
counterpart was soon identified (Orosz, Bailyn, \& Jain 1998), and a 
variable 
radio source was subsequently found at the optical position 
(Campbell-Wilson et al. 1998). These identifications were later 
confirmed by a much improved error box on the position of XTE J1550-564,
derived from an ASCA observation of the source (Marshall et al. 1998).

Shortly after the ASM discovery of XTE J1550-564, we initiated a daily 
monitoring campaign with the more sensitive detectors aboard RXTE: the 
{\it Proportional Counter Array} (PCA) and {\it High-Energy X-ray Timing 
Experiment} (HEXTE). This is a program that we have set up to study the 
X-ray properties of soft X-ray transients during the rise of an 
X-ray outburst. In this Letter, we report the discovery of strong 
aperiodic X-ray variability and quasi-periodic oscillation (QPO) in the 
X-ray light curves of XTE J1550-564, as well as the evolution of the 
observed temporal properties throughout the rising phase of the outburst.

\section{Observations}
Figure~\ref{fg:asm} shows a portion of the long-term ASM light curve of
XTE J1550-564 that highlights the rising phase of the recent outburst. 
The rising phase begins with a fast rise (with e-folding time less than 1 
day) followed by a slow rise (with e-folding time $\sim$1 week). 
During the fast rise, the soft hardness ratio appears to decrease 
while the hard ratio to increase. The X-ray spectrum softens
progressively during the slow rise, until the source settles into a 
soft state.

The times of our pointed RXTE observations are indicated in
Figure~\ref{fg:asm}. Note that the first two observations were made 
during the fast rise. For this study, we will use only the PCA 
data to investigate rapid X-ray variability of XTE J1550-564. There 
are a total of 14 observations, covering the entire rising phase of 
the outburst at a typical rate of once or twice per day. The 
effective exposure time 
varies in the range of 1.2 -- 6 ks. The first observation was a direct
spacecraft slew to the source, so it inherited the data modes ({\it
GoodXenon} with 16 s readout time) from the prior observation of 
a faint object. As a result, the data buffers were overfilled, due to 
the high count rate of XTE J1550-564, causing significant gaps in the 
{\it GoodXenon} data. These data modes were replaced, in the second 
observation, by the combination 
of a {\it Event} mode with $\sim 62\mu s$ timing resolution and a 
{\it Binned} mode with $\sim 8 ms$ timing resolution, which together 
cover the entire PCA energy band (a {\it Single Bit} mode with 
$\sim 62\mu s$ timing resolution was also used, covering the same 
energy range as the {\it Binned} mode). We subsequently optimized 
the data modes further for higher timing resolution (at the expense 
of energy resolution): a {\it Event} mode with $\sim 16\mu s$ timing 
resolution and a {\it Binned} mode with $\sim 4 ms$ timing resolution.

\section{Data Analysis and Results}
We have carried out preliminary spectral analysis. Throughout the rising 
phase, the X-ray spectrum can be roughly described by a multi-color disk 
("diskbb" in XSPEC) plus a power law. The inclusion of a gradual roll-over 
(at 16 -- 18 keV) of the power law and a Gaussian component (between 6 -- 7 
keV) significantly improves the fit. As the source brightens, the soft 
component strengthens, and the power law steepens, with the photon 
index varying from $\sim$1.3 to $\sim$2.4. Detailed spectral analysis 
(including 
the HEXTE data) will be presented in a future paper. Here, we simply use 
the preliminary model to compute {\it observed} X-ray flux (2 -- 50 keV) 
for each observation.

For timing analysis, the high-resolution data were first rebinned to 
$2^{-7}$ s, in order to facilitate comparison of results between 
observations with data modes of different timing resolution. The data
sets were then combined to cover the entire PCA energy band. A 
fast-Fourier transform (FFT) was carried out for every 256 s 
segment (with the mean subtracted and gaps filled with 0 prior to FFT) 
of each observation. Individual power density spectra were weighted and 
co-added to obtain the average power density spectrum (PDS) for that 
observation. 
The results are shown in Figure~\ref{fg:tpds}. Note that the data gaps in 
observation 1 significantly affect the low-frequency portion of the PDS, 
so what is shown (up to 4 Hz) was actually constructed from the 
{\it Standard 1} data (which is free of any data gaps). 

\begin{figure}
\psfig{figure=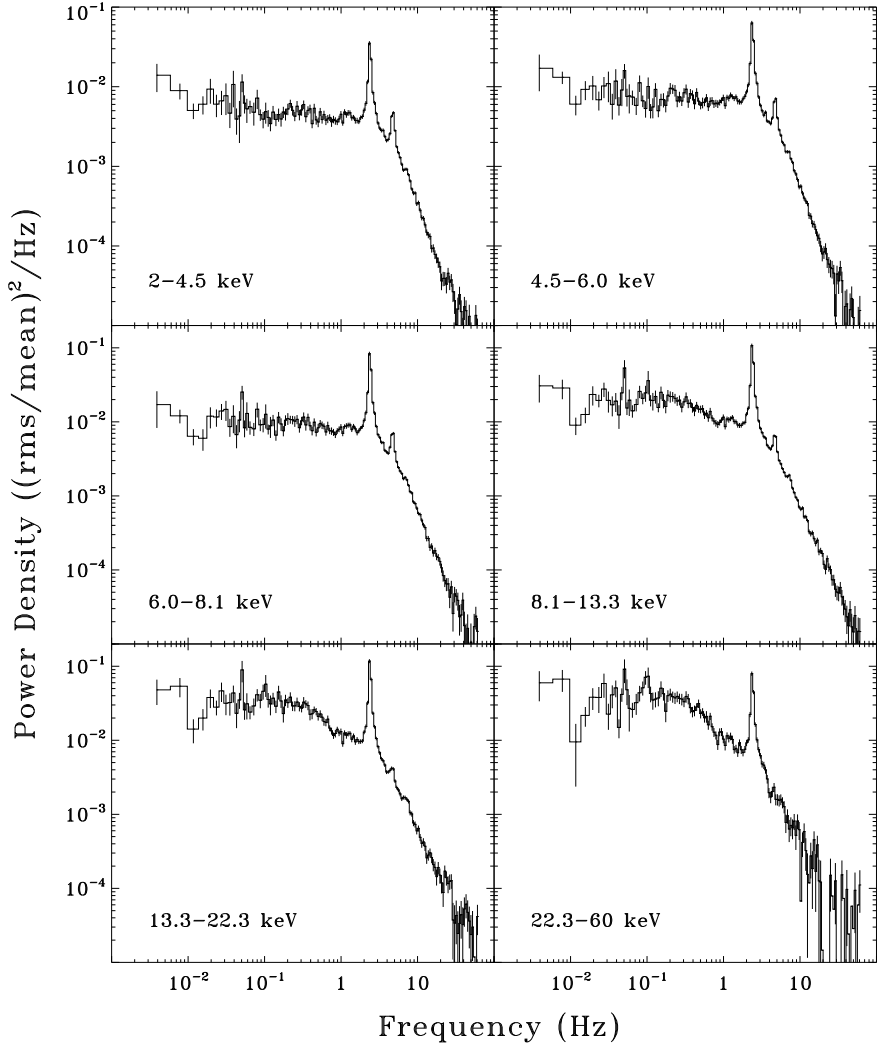,width=5in}
\caption{Power density spectra in six energy bands. This example is
for observation 8, but the results are similar for all observations. 
Note the weakening of the first harmonic of the QPO, toward high 
energies, as well as the change in the continuum shape. }
\label{fg:pds}
\end{figure}

In the first four observations, the PDS continuum is 
approximately flat at low frequencies and falls off following a power 
law at high frequencies. The presence of a pair of QPOs is apparent, 
with the higher frequency one sitting roughly at the break. There also 
exists a broad bump between 1 -- 10 Hz. We modeled the continuum as 
described, using Lorentzian functions for the QPOs and the broad
bump. The results indicate that the QPO pair are consistent with being 
harmonics of the same oscillation (Fig.~\ref{fg:qpo}). In some cases, 
the second harmonic also becomes discernible (Fig.~\ref{fg:tpds}). As 
the source brightens, the power-law portion of the PDS steepens (with 
the index dropping from $\sim -1.5$ to $\sim -1.7$). Starting from 
observation 5, the PDS can actually be better 
described by a broken power law at high frequencies, following the 
flat component. This effect becomes more apparent at higher energies 
(see Fig.~\ref{fg:pds}). The high-frequency portion of the PDS 
continues to steepen before it saturates at a power-law index of 
$\sim -2.4$ in observation 8. The break between the two power-laws 
roughly coincides with the first QPO harmonic throughout the entire
period, while the cutoff of the flat component moves little. The 
index of the shallower power-law varies in the range 
$\sim -0.8$ -- $-0.5$. Overall, the source becomes less variable (in 
terms of the fractional amplitude of the continuum;
Fig.~\ref{fg:tpds}), as it becomes brighter. 

The QPO frequency rises steeply with the increase of X-ray flux, as 
shown in Figure~\ref{fg:qpo}, roughly following a power law with index 
of +4.2 (although it seems to show sign of leveling off at high 
fluxes). While the fractional rms amplitude of the fundamental QPO 
varies little, that of the first harmonic follows a decreasing trend. 
There does not appear to be any qualitative difference in the QPO 
properties between the periods of fast rise and slow rise.

\begin{figure}
\psfig{figure=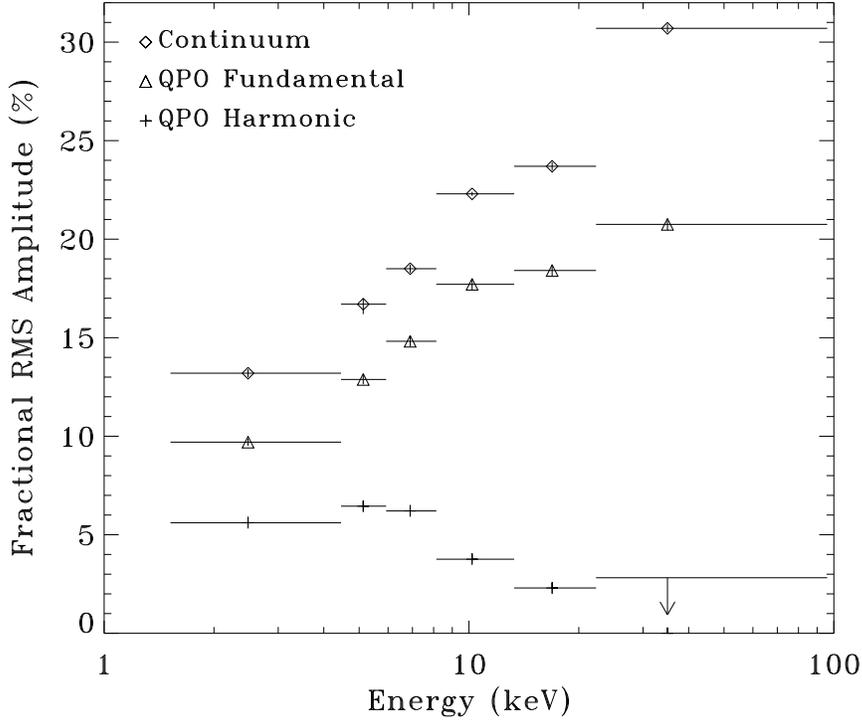,width=5in}
\caption{Energy dependence of the QPO and the PDS continuum. As in
Fig. 4, the results of observation 8 are shown to illustrate the
general dependence. The results are plotted at the effective energy 
for each band (as defined in the text). A $3\sigma$ upper limit is 
shown for the case of nondetection. }
\label{fg:param}
\end{figure}

To investigate energy dependence of the temporal properties, we 
chose six energy bands: 2 -- 4.5 keV, 4.5 -- 6.0 keV, 6.0 -- 8.1 keV, 
8.1 -- 13.3 keV, 13.3 -- 22.3 keV, and 22.3 -- 60 keV. Figure~\ref{fg:pds} 
shows the PDS in each of these bands, using observation 8 as an example 
(the results are similar for all observations). We then defined an 
``effective'' energy for each energy band,
\begin{equation}
E_{eff} = \frac{\sum_i \int E R(i,E) S(E) dE}{\sum_i \int R(i,E) S(E) dE},
\end{equation}
where S(E) is the photon flux at energy E, and R(i,E) is the detector response
matrix that distributes photons at energy E to counts in each pulse-height 
channel i; energy integrals are computed over a chosen energy band, while
all pulse-height channels are summed up. 
We fit the PDS, similarly as before, in each band.
Figure~\ref{fg:param} summarizes the results as a function of the effective 
energy. Both the continuum and the fundamental component of the QPO 
strengthens toward high 
energies, in terms of their fractional rms amplitudes, while the QPO harmonic
weakens. Our results at high energies are in general agreement with those 
from the BATSE observations (Finger, Dieters, \& Wilson 1998). 

As for the broad bump, it is initially centered at $\sim$2.4 Hz with 
FWHM $\sim$9 Hz and fractional rms amplitude 17 -- 19\% during the fast
rise (see Fig.~\ref{fg:asm}). The feature begins to weaken during the 
slow rise. The fractional rms amplitude drops to $\sim$10\% and $\sim$6\% 
in observations 3 and 4, respectively. In the mean time, the centroid 
frequency increases to $\sim$4.0 and $\sim$6.4 Hz, respectively. The
feature becomes undetectable in subsequent observations.

\section{Discussion}

The rise of XTE J1550-564 shows a more complicated profile than the 
canonical exponential rise (see Chen, Shrader, \& Livio 1997 for a 
compilation of 
outburst profiles). During the fast rise, the source brightened at 
nearly all wavelengths (soft and hard X-rays, optical, and radio). 
This is consistent with a sudden surge in the mass accretion rate at 
onset of the outburst, as generally thought (see review by King 1995).
During the slow rise, however, the ASM flux (1.3 -- 12 keV) becomes 
{\it anti-correlated} with the BATSE flux (25 -- 200 keV),\footnote{see 
http://www.batse.msfc.nasa.gov/data/occult/fluxhistories for BATSE
measurements} similar to Cyg X-1 during its transition between the 
hard and soft states (Cui et al. 1997a; Zhang et al. 1997). It is 
interesting to note that the soft and hard fluxes became 
{\it positively} correlated again during a giant 
flare around MJD 51076. In fact, the BATSE flux reached nearly the same 
level during the flare as that at the end of the fast rise. Perhaps
the same physical processes operate in initiating the outburst and in 
setting off the flare. 

\begin{figure}
\psfig{figure=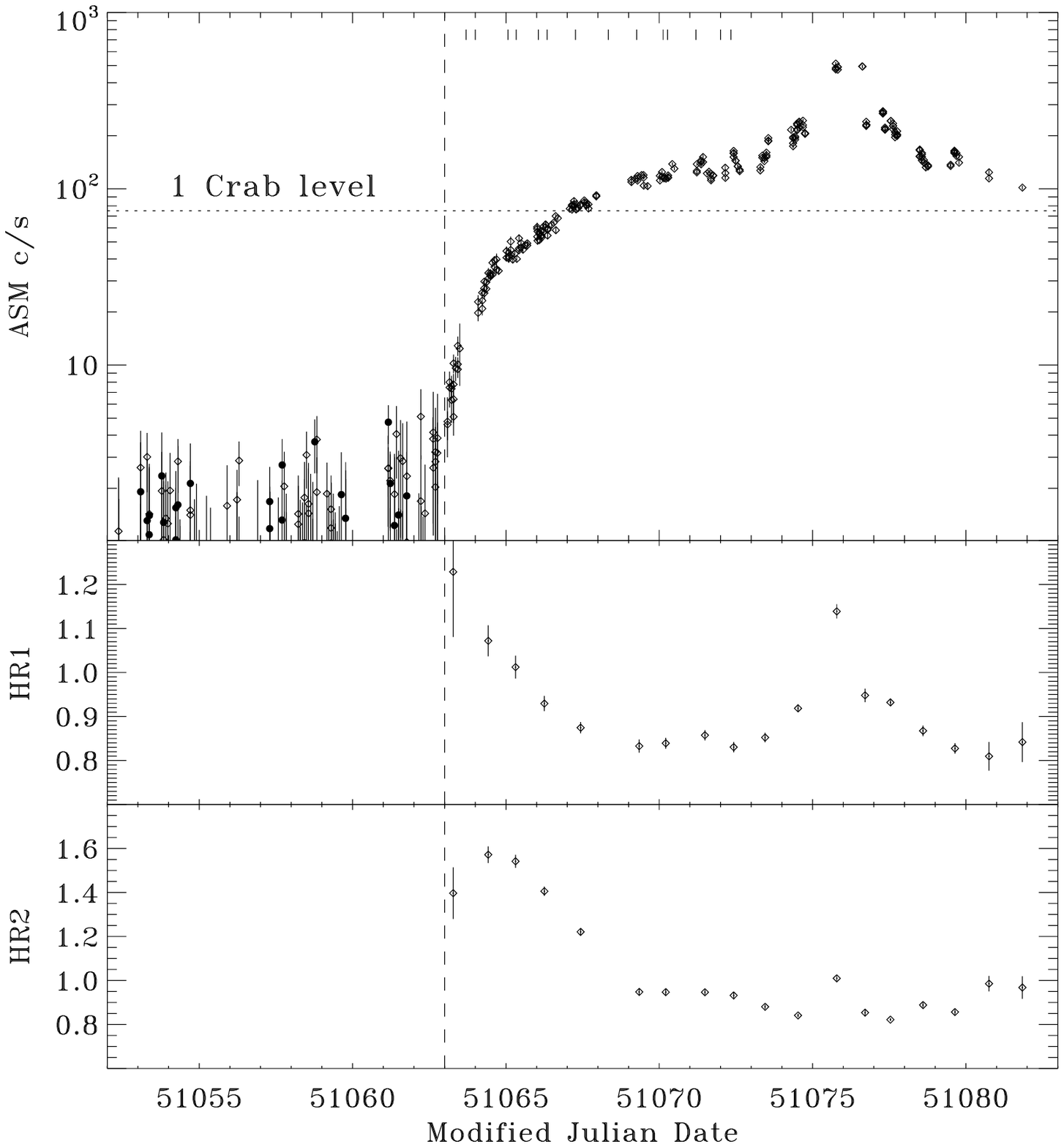,width=4in}
\caption{ASM light curves of XTE J1550-564. {\it Top:} Each data 
point is derived from a 90 s measurement, with negative values 
shown in filled circles for completeness. The dashed line indicates 
roughly the time (MJD 51063 or 9/7/98 UT) when the outburst started. 
For reference, the dotted line shows the ASM count rate of the Crab
Nebula. The short vertical lines at the top indicate the times of 
the pointed RXTE observations reported in this Letter. {\it Middle:} 
Soft hardness ratio is defined as the radio of the count rate in the
3.0 -- 4.8 keV band to that in the 1.3 -- 3.0 keV band. Shown are the
daily-averaged results. {\it Bottom:} Hard hardness ratio is defined 
as the radio of the count rate in the 4.8 -- 12 keV band to that in the 
3.0 -- 4.8 keV band. }
\label{fg:asm}
\end{figure}

The observed rapid X-ray variability and hard energy spectrum make 
XTE J1550-564 a likely black hole candidate (BHC). Prior to the launch 
of RXTE, QPOs in BHCs 
were observed only at very low frequencies (less than 1 Hz, except for the 
``very-high-state'' QPOs at a few Hz observed only on two occasions; 
see van der Klis 1995). RXTE has pushed the 
discovery space up to hundreds of Hz (see Cui 1999). The PDS of XTE 
J1550-564 bears resemblance to that of GRS 1915+105 in the ``low-hard'' 
state (Morgan, Remillard, \& Greiner 1997), where a strong QPO was 
detected, as well 
as its first harmonic in some cases. Following Morgan et al. (1997), 
we have tracked individual pulses of the QPO (fundamental component) 
in XTE J1550-564. The results imply that the QPO represents a random 
walk in the oscillation phase, like its counterpart in GRS 1915+105. 
However, while the QPO frequency varies with X-ray flux for both 
sources, the exact frequency-flux correlation is different 
{\it quantitatively} (comparing our results with those of Chen, Swank, 
\& Taam 1997). In general, the presence 
of QPOs seems to be characteristic of BHCs during the transitions 
between the hard and soft states (Cui 1999). Therefore, the slow 
rise of XTE J1550-564 may indeed represent such a transition, as also 
indicated by the observed spectral evolution (see Fig.~\ref{fg:asm}). 
On the other hand, the canonical ``1/f'' PDS shape for the soft state 
(van der Klis 1995; Cui 1999) was not seen here.

\begin{figure}
\psfig{figure=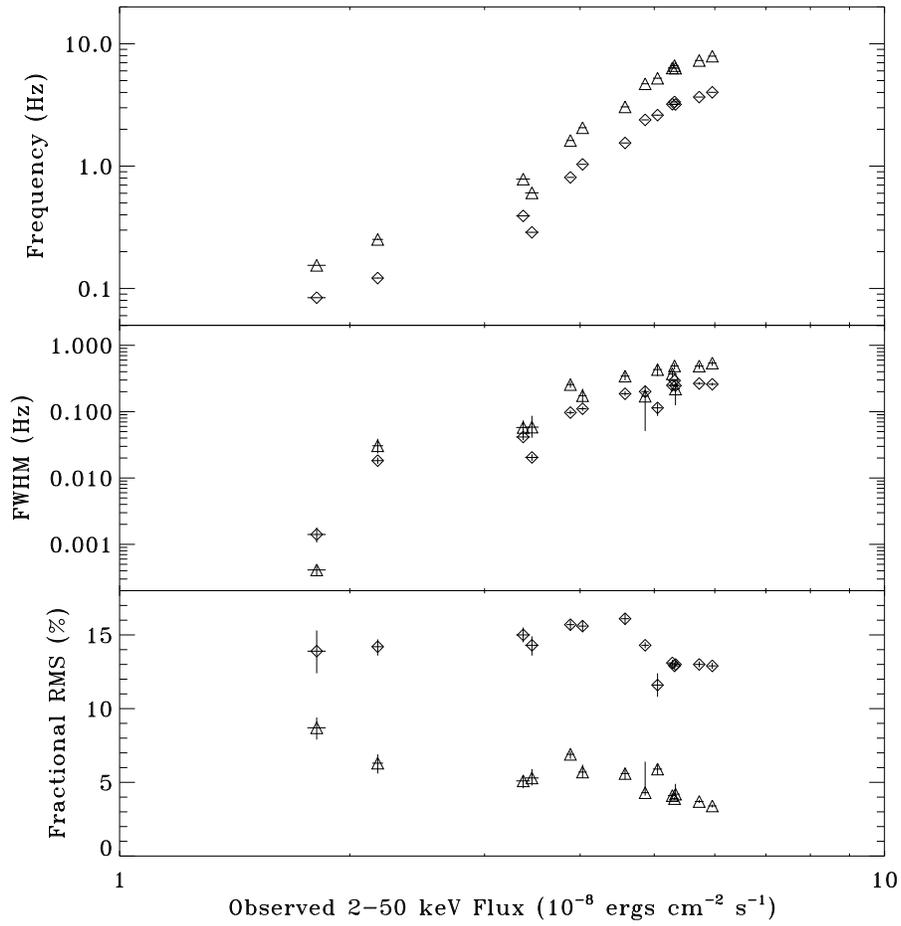,width=5in}
\caption{Evolution of the QPO during the rising phase of the outburst. 
Diamonds show the fundamental component, and triangles show the first 
harmonic. }
\label{fg:qpo}
\end{figure}

XTE J1550-564 has offered the best case of tracking a QPO over a wide 
range of X-ray flux. Previous studies were usually hindered by the 
ambiguity in ``labeling'' a moving QPO as the X-ray flux varies, due 
to insufficient coverage of the phenomenon. After all, QPOs are often 
transient phenomena; different QPOs can be present in different 
observations for a particular source. Here, XTE J1550-564 was
monitored at least once a day, so the evolution of the QPO could 
be closely followed. The presence of the strong harmonic component also 
helps uniquely identify the feature. Therefore, the observed 
correlation between the QPO properties and X-ray flux is reliable. 
Such correlation seems to suggest a different origin for the QPOs in 
XTE J1550-564 than the 67 Hz or 300 Hz QPOs in microquasars whose 
frequency is stable against any variation in the X-ray flux and than 
those (in BHCs in general) whose frequency decreases with the flux
increase. It is worth commenting that the reported weak QPO at 184 Hz 
during the giant flare (McClintock et al. 1998) is probably not of the 
same type, since an extrapolation of the best-fit power law of the
frequency-flux relation (Fig.~\ref{fg:qpo}) would require a flux 
increase only 
by a factor of $\sim$2.5 during the flare. This is much less than the 
observed increase (by a factor of $\sim$3.9 in the ASM count rate, 
thus, more in the flux because the energy spectrum hardens during the 
flare). Moreover, although the QPO seen here broadens as the source 
becomes brighter, the Q-value ($\equiv f_{qpo}/FWHM$) tends to
increase mildly, reaching $\sim$15 (derived from Fig.~\ref{fg:qpo}), 
which is much larger than that of the 184 Hz QPO ($\sim$4).

\begin{figure}
\psfig{figure=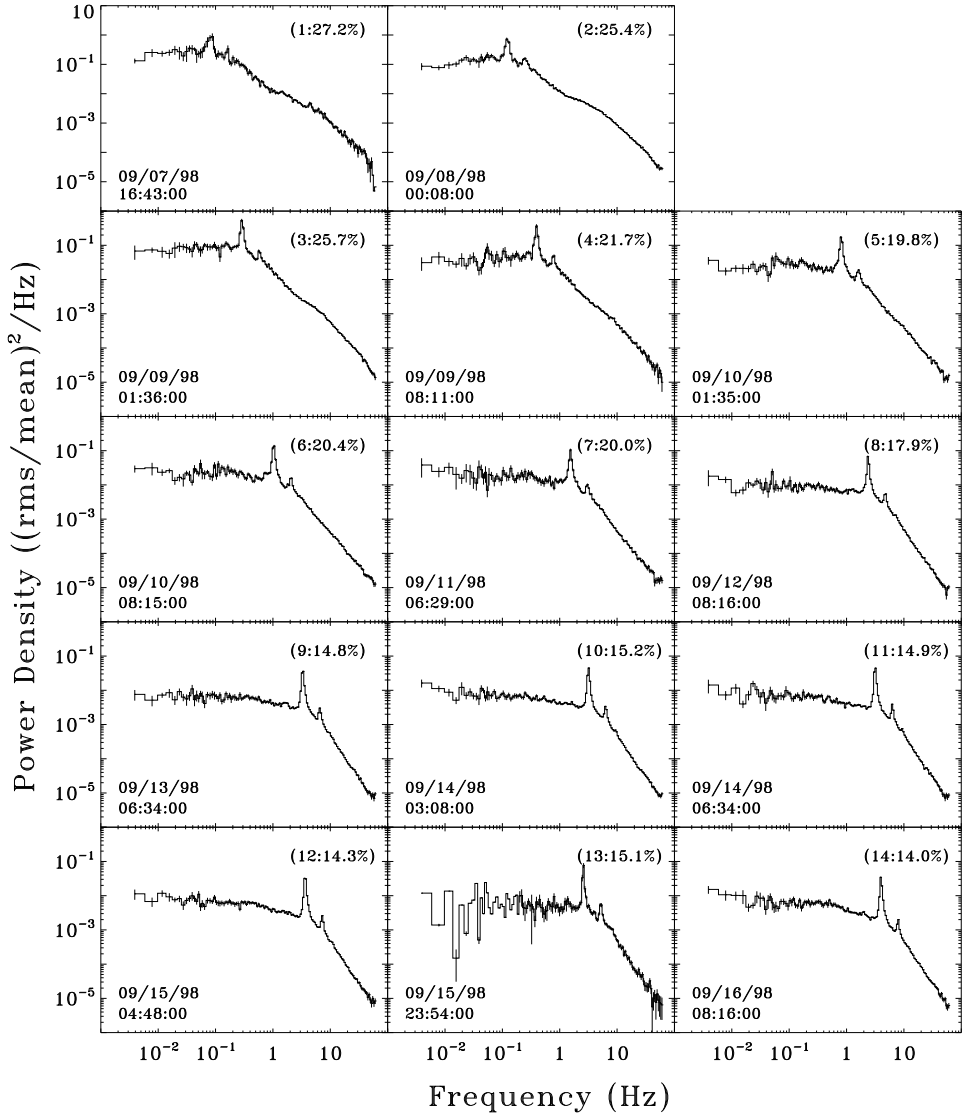,width=5in}
\caption{Power density spectrum. The start time (in UT) of each 
observation is shown, as is the overall fractional rms amplitude of 
the continuum over the frequency range 0.001 -- 64 Hz.}
\label{fg:tpds}
\end{figure}

Like most QPOs observed in BHCs (Cui 1999, and references therein), the
fundamental component of the QPO in XTE J1550-564 strengthens toward 
high energies. Interestingly, however, the QPO harmonic becomes weaker, 
which is again similar to the low hard state QPO in GRS 1915+105, as
well as to the very high state QPOs (Belloni et al. 1997; Takizawa et
al. 1997). The energy dependence of QPO amplitudes has not been
emphasized enough but might hold the key to our understanding of the 
origin of these QPOs (Cui 1999). The observed dependence seems to rule 
out the possibility that these QPOs are produced by any oscillatory 
processes in a standard thin accretion disk (Shakura \& Sunyaev 1973), 
since the X-ray emission from such a disk is very soft (less than a
few keV). The hard X-ray emission is generally thought to be the
product of energetic 
electrons (thermal or non-thermal) inverse-Compton scattering soft 
photons in the system. The scattering electrons then serve as a 
``low-pass filter'' to any intrinsic variability associated with the 
seed photons (Cui et al. 1997a; Kazanas, Hua, \& Titarchuk 1997). 
Therefore, the 
increase of the PDS break frequency as the source brightens (see 
Fig.~\ref{fg:tpds}) perhaps implies a decrease in the size of the 
Comptonizing region, as is being suggested for Cyg X-1 during the state 
transition (Cui et al. 1997a, 1997b). The QPO might simply 
originate in the Comptonizing region, due to oscillation in the
density and/or energy distribution of hot electrons during the rising
phase of the outburst (Cui et al. 1997b; Kazanas et al. 1997; 
Titarchuk, Lapidus, \& Muslimov 1998). It remains to be seen whether 
the observed 
properties, such as dependence of the QPO properties both on photon 
energy and X-ray flux (i.e., mass accretion rate), can be accounted for 
{\it quantitatively} in this scenario. The origins of QPOs in BHCs are 
still poorly understood at present (van der Klis 1995; Cui 1999). Few 
detailed models exist, due to the lack of data. Hopefully, more and 
better data provided by missions like RXTE will soon lead to quantitative 
modeling of the phenomena.

\acknowledgments
We thank Dr. Michiel van der Klis for useful comments. This work is 
supported in part by NASA grants NAS5-30612 and NAG5-7484.

\clearpage

\clearpage

\clearpage

\clearpage

\clearpage

\end{document}